\documentclass[traditabstract]{aa}
\usepackage{graphicx}
\usepackage{setspace}
\usepackage{longtable,lscape}
\usepackage{hhline}
\usepackage{array}
\usepackage{txfonts}
\title{\textbf{Search for associations containing young stars (SACY):}}
\subtitle{II. Chemical abundances of stars in 11 young Associations in the Solar neighborhood
\thanks{Based on observations collected with the UVES spectrograph at the 
VLT/UT2 8.2-m Kueyen Telescope (ESO run ID. 079.C-0556(A)) at the Paranal Observatory, Chile}}
\author{P. Viana Almeida
          \inst{1,2,6}
	\and
          N.C. Santos
	  \inst{1,6}
	\and
          C. Melo
	  \inst{3}
	\and
	  M. Ammler-von Eiff
		\inst{4}
	\and
	  C. A. O. Torres
		\inst{5}
	\and
	  G.R. Quast
		\inst{5}
	\and
	  J.F. Gameiro
		\inst{1,6}
	\and
	  M. Sterzik
		\inst{2}
          }
\institute{Centro de Astrof{\'i}sica, Universidade do Porto, Rua das Estrelas, 4150-762 Porto, Portugal
         \and
             ESO, Alonso de Cordova 3107, Casilla 19001, Vitacura, Santiago, Chile\\
             \email{palmeida@eso.org}
	  \and
             ESO, Karl-Schwarzschild-Str.2, 85748 Garching bei M\"{u}nchen, Germany
	  \and
             Centro de Astronomia e Astrof{\'i}sica da Universidade de Lisboa, Observat\'{o}rio Astron\'{o}mico de Lisboa, Tapada da Ajuda, 
	     1349-018 Lisboa, Portugal
	  \and
	     Laborat\'{o}rio Nacional de Astrof{\'i}́sica/ MCT, Rua Estados Unidos 154, 37504-364 Itajub\'{a}́ (MG), Brazil
	  \and
	     Departamento de Matem\'{a}tica Aplicada, Faculdade de Ci\^{e}ncias da Universidade do Porto, Portugal
		   }
  \abstract
   {The recently discovered coeval, moving groups of young stellar objects in the solar neighborhood 
   represent invaluable laboratories to study recent star formation and to search for high metallicity stars 
   which can be included in future exo-planet surveys.
   In this study we derived through an uniform and homogeneous method stellar atmospheric parameters and abundances for iron, silicium and 
   nickel in 63 Post T-Tauri Stars from 11 nearby young associations. We further compare the results with two different 
   pre-main sequence (PMS) and main sequence (MS) star populations. The stellar atmospheric parameters and the abundances presented here were derived 
   using the equivalent width of individual lines in the stellar spectra through the excitation/ionization equilibrium of iron. 
   Moreover, we compared the observed Balmer lines with synthetic profiles calculated for model atmospheres with a different line formation   
   code. We found that the synthetic profiles agree reasonably well with the observed profiles, although the Balmer lines of 
   many stars are substantially filled-in, probably by chromospheric emission. Solar metallicity is found to be a common trend on all the 
   nearby young associations studied. 
   The low abundance dispersion within each association strengthens the idea that the origin of these nearby young associations is 
   related to the nearby Star Forming regions (SFR).
   Abundances of elements other than iron are consistent with previous 
   results on Main Sequence stars in the solar neighborhood. The chemical characterization of the members of the newly found nearby young 
   associations, performed in this study and intended to proceed in subsequent works, is essential for understanding and testing the context 
   of local star formation and the evolutionary history of the galaxy.}

\begin{document}
\maketitle
\section{INTRODUCTION}
For many years the canonical star-forming regions  (such as Taurus, Rho-Oph, Sco-Cen complex, for instance) were
the closest places to study astrophysical issues related to the (local and recent) star-formation process.\\
\indent The confirmation by Kastner et al. (\cite{kas97})
of an early suggestion made by Gregorio-Hetem (\cite{gregorio92})
that a small group of young stars around TW Hydra would form a loose
association sharing common characteristics such
as age, distance, x-ray emission level and radial velocities, 
showed that young stars can be found at a much closer distance than previously thought.\\
\indent Since then, a dozen of young nearby associations have been identified
 (see reviews by Zuckermann \& Song \cite{zucksong}; Torres et al. \cite{tor03};
Torres et al. \cite{tor08}). These associations are valuable laboratories
to study recent star formation in the solar neighborhood. \\
\indent Thanks to the proximity, they may
yield precious clues that can lead to a better understanding of a wide variety of current open issues
concerning the formation and evolution of planetary systems.\\
%
%
\indent Because typical ages of these groups match the timescales for disk dispersal (\textit{e.g.} 
Spangler et al. \cite{spa01}) and planet formation (Pollack et al.~\cite{pol96}) 
they constitute interesting targets for exoplanet searches. \\
\indent As previous studies strongly suggest (\textit{e.g.} Gonzalez
\cite{go97}; Gonzalez et al. \cite{go01}; Santos et al. \cite{sa01}; Santos et al. \cite{sa04b};  Santos et al.
\cite{santo05}; Fisher \& Valenti \cite{fixe05}), planet-host stars may display a higher metal content.
Consequently it becomes relevant to look for a high metallicity sample among young stars, such as these Post-T Tauri stars (PTTS), 
in order to better focus future planetary search projects.\\
\indent In this context, it is important to carry out a chemical characterization
of the newly-found loose associations and compare them with the typical
stellar population, both the older MS (Sousa et al. \cite{sousinha}) and the younger PMS (James et al.
\cite{ja06}; Santos et al. \cite{sa08}) found near the Sun. That comparison could lead to a better understanding
of the origin of the stellar population in the solar neighborhood.\\
\indent In this study we present stellar parameters and chemical abundances for 63 PTTS in 11 young nearby
associations, being them the AB~Doradus (AB Dor) (Zuckermann et al. \cite{zusobe04}; Torres et al.
\cite{tor08}), Argus (Torres et al. \cite{tor08}), $\beta$ Pictoris (Zuckermann et al. \cite{zucka01}; Song et
al. \cite{songa03}; Torres et al. \cite{tor06}; Torres et al. \cite{tor08}), Carina , Columba (Torres et al.
\cite{tor08}), $\varepsilon$ Chamaeleon (Barrado y Navascu{\'e}s \cite{barna98}; Mamajek et al. \cite{mamaj00};
Torres et al. \cite{tor08}), R~Coronae Australis association  (R CrA) (Neuh{\"a}user et al. \cite{neu00}), Lower
Centaurus-Crux (LCC), Octans (Torres et al. \cite{tor08}), Tucana-Horologium (Tuc-Hor) (Zuckermann $\&$ Webb
\cite{zuwe00}; Torres et al. \cite{to00}; Torres et al. \cite{tor08}) and Upper Scorpius (US). Furthermore, we
compare our results with the ones provided by a sample of PMS and MS population in the solar neighborhood taken
from the literature. A subsequent discussion on the implications of the study is presented.\\
\section{Sample and observations}
The 63 objects used in this study are listed in Table~\ref{radec}. 
They were selected from  Torres et al. (\cite{tor06}) based on
the following criteria: i) they have been classified as highly probable or probable member of a given
association according to 
Torres et al. (\cite{tor08});
ii) they  show $\upsilon\sin i$ less than 15 km/s in order to avoid strong line blending; 
iii) to the best of our knowledge, they are all single stars;
and iv) they are young (younger or at Pleiades age) based on the equivalent width of Li I 6708 line.\\
\indent High-resolution (R = $\frac{\lambda}{\Delta\lambda}$ = 50.000 in both arms) spectroscopic observations carried
out between March and September 2007 were performed using the DIC\#1 390+580 mode with the UVES spectrograph (Dekker et al.\cite{dek00}) at the VLT/UT2 8.2-m Kueyen Telescope (Paranal Observatory, ESO, Chile). The spectra obtained cover a wavelength interval from 4800 to
7000 \AA{} with a gap between $\sim$ 5755 and 5833 \AA{}.\\
\indent Data reduction was performed with the UVES pipeline by carrying out the usual steps for echelle data reduction. 
For most of the targets, multiple exposures had to be taken in order to avoid saturation. Once 
extracted, the multiple spectra for a given target were put at rest velocity and combined to make a high SNR ($\sim$150-200) final spectra.\\
\indent The synthesis of the H$\alpha$ and the H$\beta$ profiles was carried out using the 2D extracted spectra (i.e., the
intermediate pipeline product prior to the merging of the spectral orders). 
This step is necessary  in order to properly normalize the regions of
the Balmer profiles to the relative continuum. We refer the reader to the work of Ammler-von Eiff \& Santos
(\cite{ammsant08}) for a detailed explanation of the different steps of the reduction procedure. \\
\indent The comparison data for MS dwarfs used in this paper were taken from 
Sousa et al. (\cite{sousinha}) (hereafter SOU08) and Gilli et al. (\cite{gi06})
(hereafter GI06) whereas the WTTS comparison sample comes from Santos et al. (\cite{sa08}) 
(hereafter SA08). 
\section{Analysis of Spectra}
\subsection{Stellar parameters and abundances}
The stellar parameters and iron abundances shown here were determined using the same methodology described in Santos el al.
(\cite{sa04b}). Preliminary work consisted in measuring the equivalent widths (EW) of a set of FeI and FeII lines on the
collected spectra using the IRAF splot tool and a gaussian fitting procedure. The line list of the FeI/FeII species
considered was taken from aforementioned paper and is composed essentially by weak lines with an extended range of excitation
potentials in order to ensure a robust fitting. \\
\indent A spectroscopic analysis was then performed with a recent version (2002) of the radiative transfer code MOOG
(Sneden et al. \cite{sne73}) using the local thermodynamic equilibrium (LTE), constant flux, blanketed atmospheric models,
taken from the ATLAS9 stellar grid of Kurucz (\cite{kur93}).\\
\begin{figure}[t]
\centering
\includegraphics[scale=0.45]{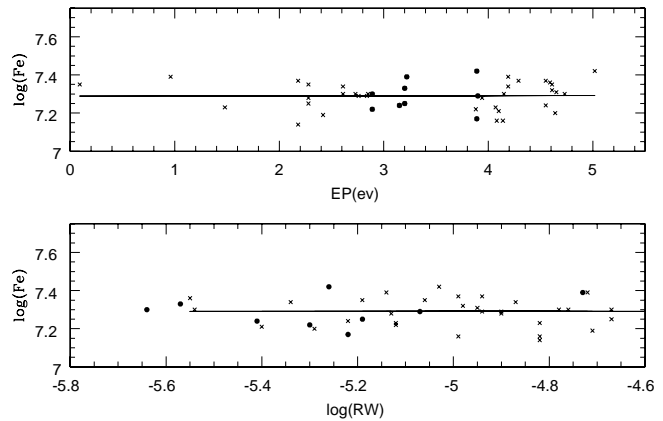}
\vspace*{-2.8cm}
\caption{Individual line FeI (crosses) and FeII (filled circles) abundances as a function of the excitation potential (top panel) and the reduced equivalent width, RW = EW/$\lambda$ (bottom panel) for one of the coolest stars in the sample, CD-22 11502 in the US association. In both cases, no systematic dependence of the derived abundances is seen as a function of these two quantities.}
\label{logrw}
\end{figure}
\indent The four fundamental parameters, effective temperature (T$_{eff}$), surface gravity (log$g$), microturbulence
($\xi_{t}$) and the Iron abundance ([Fe/H]) presented here were furthermore derived by imposing the Fe excitation/
ionization equilibrium. \\
\indent In Fig.~\ref{logrw} we show an example of the two final fits (slope $\sim$ 0.00) obtained for a cool star (CD-22 11502) with a T$_{eff}$ of 4951K. Of particular interest in those plots is the small dispersion of the iron abundances found.\\
\indent The errors affecting the parameters were computed through the very same method as in Gonzalez
(\cite{gon98}). Final derived parameters, abundances, and uncertainties are presented in Table~\ref{kstars}. We should point out
that the errors displayed in this study only represent relative uncertainties. They do not assert absolute uncertainties on the atmospheric parameters but rather underlying errors stacked to the method used in this study. \\ 
\indent In addition to Iron, abundances for Si and Ni were also derived. We refer the reader to the papers of
Santos et al. (\cite{sant06}) and GI06 for a description of the
procedures and atomic line list used. Resulting abundances
and errors for these elements are given in Table~\ref{elements}.
Furthermore, abundance ratios [X/Fe] (with X = Ni and Si) were compared with those
from GI06, who applied the same method, line list and atmospheric models.\\
\indent All  abundances displayed here are differential with respect to the Sun. The following solar parameters were
adopted: T$_{eff}$ = 5777 K, log$g$ = 4.44 dex, $\xi_t$ = 1.00 km s$^{-1}$, log$\varepsilon_{Fe}$ = 7.47 dex. 
Solar abundances for these for Si and Ni were taken from Anders \& Grevesse (\cite{ag89}).\\
\indent Similarly to previous studies on PMS stars (Padgett \cite{padgee96}; James et al. \cite{ja06}; Santos et al.
\cite{sa08}), the microturbulence values obtained for the PTTS studied here are somewhat higher than the typical ones
usually exhibited by MS field dwarfs. The reason for these larger values is unknown. It has been suggested
that this may be caused by an enhanced magnetic field activity known to be considerably stronger during 
the pre-main sequence phase (SA08). \\
\indent We plotted the microturbulence against the various parameters involved in the determination, such as the T$_{eff}$,
log$g$, [Fe/H], and $\upsilon$ sin$i$ in order to detect possible trends which could be at the origin of these higher
microturbulence values.
Despite a smooth dependence on the temperature, the cooler the star the higher the $\xi_t$, it seems that there is no
apparent influence of other factors such as the $\upsilon$ sin$i$ of the stars. In this context, we were not able to
understand whether these results come from the intrinsic physics of the stars studied here or, for instance, from the
atmospheric models used.\\
\addtocounter{table}{2} 
\setlength\extrarowheight{2pt}\vspace*{0.5pt}
\begin{table}[bt]
\caption{Average Fe abundances and standard deviations found in each association. If only one star is studied the $\sigma$ displayed comes from the determination of its [Fe/H]. The average values of [Fe/H] found in the stellar sample of SA08 and SOU08 are also shown. The columns marked with an asterisk are the corrected average [Fe/H] and the new $\sigma$ computed according to the method presented in section 3.2.}
\label{table:1}      
\centering                          
\begin{tabular}{cccccc}        
\hline\hline                 
Association    & $\langle$[Fe/H]$\rangle$    &   $\sigma$  & $\langle$[Fe/H]$\rangle$* & $\sigma$* & N$_{stars}$ \\  
\hline                                                 
AB Dor     &	-0.01	   &	0.09  & 0.04 & 0.05 & 12\\ 
Argus     &	-0.03	   &	0.05  & 0.02 & 0.06 & 7\\
$\beta$ Pic &	-0.13	   &	0.08  & -0.01 & 0.08 & 1\\ 
Carina	  &	-0.07	   &	0.04  & 0.01 & 0.04 & 7\\ 
Columba	  &	-0.05	   &	0.07  & -0.03 & 0.07 & 6\\
R CrA	  &	-0.08	   &	0.00 & 0.00 & 0.00 & 2 \\
$\epsilon$ Cha	  & 0.01   &	0.05 & 0.03 & 0.05 & 1\\
LCC	  &	-0.06	   &	0.05  & 0.02 & 0.05 & 7\\	 
Octans  & 	-0.09	   &	0.06  & -0.03 & 0.05 & 1\\
Tuc-Hor	  & 	-0.06	   &	0.09 & -0.03 & 0.05 & 9 \\	 
US	  & 	-0.11	   &	0.10  & -0.02 & 0.09 & 10\\
\hline 
Total sample & -0.06 & 0.08  & 0.00 & 0.06 & 63\\
\hline\hline
SA08 & -0.08 & 0.09 & 0.04 & 0.06 & 28\\
SOU08 & -0.10 & 0.24 & - & - & 450\\
\hline 
\end{tabular}
\end{table}
\subsection{[Fe/H] correction}
\begin{figure}[t]
\centering
\includegraphics[scale=0.45]{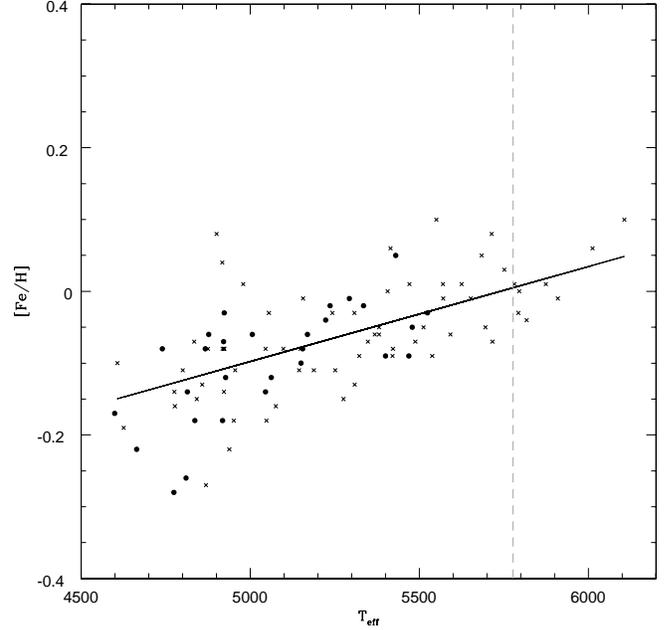}
\caption{T$_{eff}$ vs [Fe/H] plot for the stars studied in this work (crosses) and that of SA08 (filled circles). The solid black line represents a simple least square fit of the initial values found. The slope of the line is of about 0.13dex \textit{per} 1000K. The dashed vertical line represents the solar value used as a reference.}
\label{fehmod}
\end{figure}

Inspecting the atmospheric parameters and Fe abundances shown in Table~\ref{kstars} we note that in the
temperature range from 4500K to 5500K the determined values of [Fe/H] suffer an overall increase as a function of T$_{eff}$
(Fig.~\ref{fehmod}).
This fact is true for any given association studied.
Overplotting the [Fe/H] and T$_{eff}$ derived by SA08 to our own values we promptly see that the trend
reported by these authors is similar to the one found in our analysis.\\
\indent The trend shown in Figure~\ref{fehmod} contradicts the standard idea that stars originated within the same cloud
will show a similar abundance pattern. 
Therefore, we believe that the slope seen in Fig.~\ref{fehmod} is likely to be due to systematic errors. 
In this respect, it is known that PMS stars often display an enhanced stellar activity and the spectroscopic
results, like atmospheric parameters, may become modulated by stellar spots, strong magnetic fields or NLTE effects. 
Moreover, as the star gets cooler, the line blending becomes stronger, 
causing  a possible [Fe/H] underestimation.
An extensive study of the influence of stellar spots on deriving stellar parameters can be found, for instance, in Morel et
al. (\cite{morel04}).\\
%
%
\indent A possible influence of the atmospheric models used, optimal in the temperature window considered, cannot be
discarded. However, tests carried out in Sec. 3.3 (see below) do not support such hypothesis. A robust comparison
sample with different stellar models is needed in order to understand if the trend observed is a by-product of the analysis
employed or if it has a real origin in the stars studied. \\
\indent Regardless the origin of this effect,
a first-order correction was carried out by performing  a simple linear
least square fit to the data points in Fig~\ref{fehmod}. We then subtracted from all [Fe/H] values
the fitted relation, assuming that for the solar temperature no correction was needed. 
The same correction was applied for the
SA08 metallicities.
The [Fe/H] values corrected by this method are shown in the asterisked 
columns in Table~\ref{kstars} and Table~\ref{table:1}.\\
\indent By looking at the corrected results in these two tables we see that a slightly smaller dispersion within 
each association is found. However, since we cannot find a physical justification for the correction applied above, only 
the original (uncorrected) values of [Fe/H] obtained were considered throughout the paper.
\subsection{H$\alpha$ and H$\beta$ synthetic profiles}
\indent To further establish the consistency of the stellar parameters, and more specifically of T$_{eff}$ as determined using the method described in section
3.1, we computed the synthetic profiles for the two Balmer lines, H$\alpha$ and H$\beta$. It has been shown that H-line profiles are very sensitive to temperature variations, but quite insensitive to gravity and metallicity variations (see e.g. Fuhrmann et al. \cite{frufru04} and references therein). That is the reason why it is possible to derive quite accurate stellar T$_{eff}$ from the hydrogen lines.\\
\indent Ten stars of the whole sample were used, half of them featured
in our analysis as having T$_{eff}$ lower than the Sun and the other half with approximately solar T$_{eff}$. \\
\begin{figure}[t]
\centering
\includegraphics[scale=0.55]{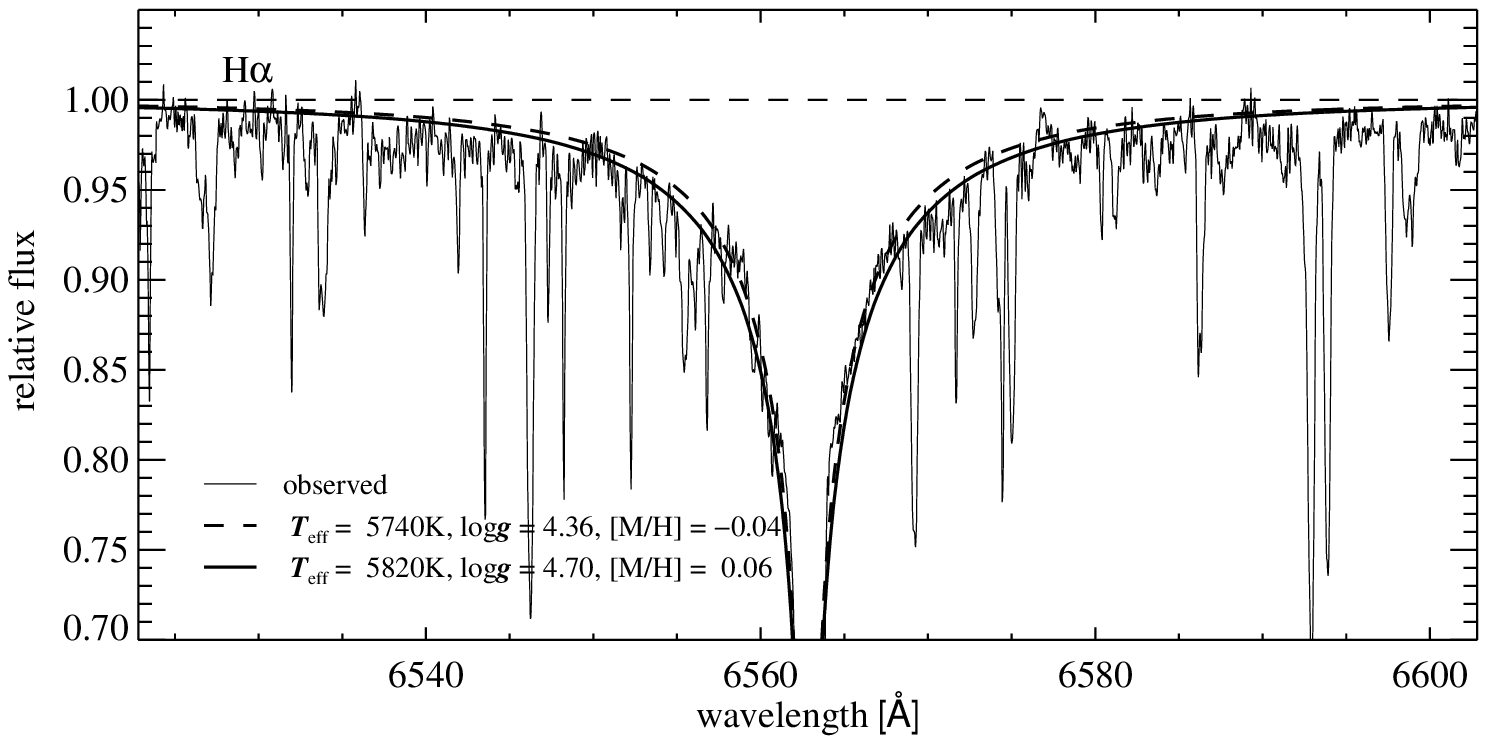}
\includegraphics[scale=0.55]{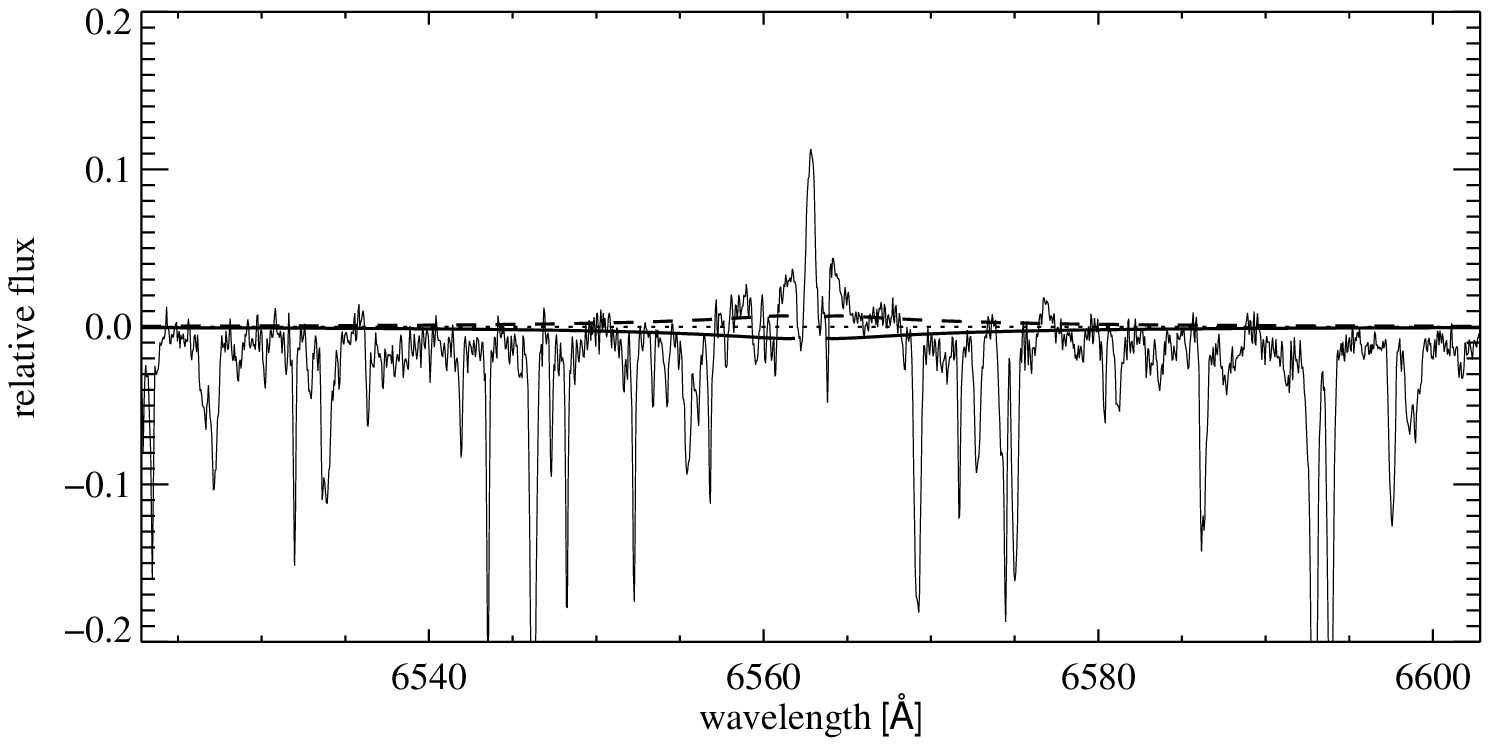}
\caption{Observed and synthetic H$\alpha$ profile for the star HD47875. The horizontal dashed line represents the assumed continuum. One profile was computed for a lower limit (dashed line) on all parameters and another for an upper limit (solid line). The residuals of the yielded comparison are displayed in the lower panel.}
\label{halpha}
\end{figure}
\indent The profiles were calculated with the LINFOR line formation code using the MAFAGS model atmospheres as described in
Fuhrmann et al. (\cite{frufru04} and previous papers), Ammler (\cite{ammler06}), Ammler-von Eiff \& Santos (\cite{ammsant08}).
We then compared the synthetic profiles with those observed in the stellar spectra. The method used consisted in picking the
T$_{eff}$, log$g$, $\xi_{t}$ and the [Fe/H] in Table~\ref{kstars} to derive two profiles for both the H$\alpha$ and H$\beta$
lines for each star in the sample. One profile is calculated for the lower limits on all parameters and one for the upper limits
on all parameters given by the uncertainty range on the parameters in Table~\ref{kstars}. This is a good choice since reducing
log$g$ and [Fe/H] has the same effect on the profiles as reducing T$_{eff}$, which maximizes the effect of the uncertainties
onto the synthetic profiles. \\
\indent Note that the comparison of the synthetic profiles to the observed profiles is only valid down to relative flux levels
of about 70 percent. Below this value, deviation due to NLTE starts to play a role Fuhrmann et al. (\cite{frufru04} and previous
papers).\\
\indent For some stars, the H$\alpha$ lines seemed strongly affected by chromospheric emission, the strongest displayed by
CD-361785. The H$\beta$ seems less affected throughout the whole sample. We decided to probe the influence of this emission on
the Balmer line wings and implications for derivation of effective temperature by computing the residuals as displayed in the
lower panel of Fig~\ref{halpha}. \\
\indent In this plot, the limiting synthetic profiles were not subtracted from the observed profile. Instead, an additional
synthetic profile was calculated based on the nominal stellar parameters (regardless of the error bars). This synthetic profile
then was subtracted from the limiting synthetic profiles as well as from the observed profile. These differences are shown in
the lower panel of Fig~\ref{halpha}. \\
\indent By studying these residuals it turned out that for the chromospherically active stars H$\beta$ seemed a  better
choice for confirming effective temperature. However there are usually  many lines in the H$\beta$ region and the wings
cannot be properly traced. Still H$\beta$ can be used to confirm T$_{eff}$ when H$\alpha$ becomes useless, e.g. in the example
of CD-361785.\\
\begin{figure}[t]
\centering
\includegraphics[scale=0.45]{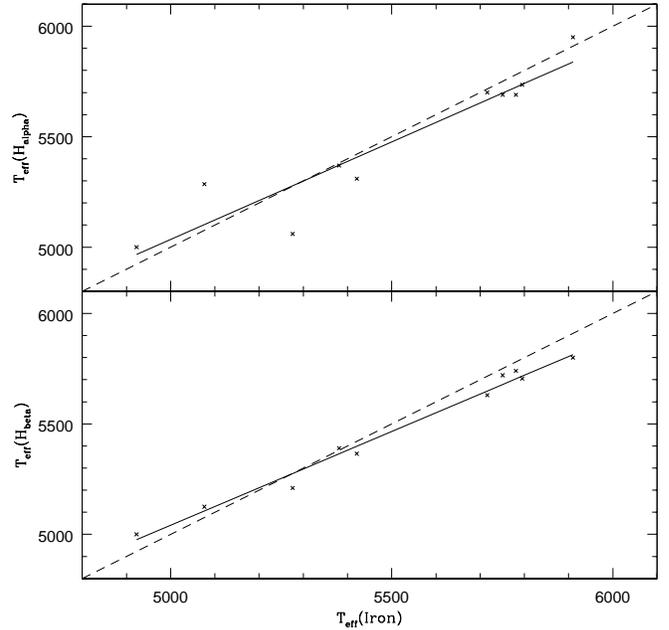}
\vspace*{-0.7 cm}
\caption{Comparison between effective temperatures derived using iron lines
with those obtained through synthesis of the
line profiles of H$\alpha$ and H$\beta$. Both solid lines represent linear
least square fits performed to the data (with a slope of $\sim$ 0.9).
For clarity we also show the ideal 1:1 relation (dashed line).}
\label{halpha_beta}
\end{figure}
\indent In the two panels of Fig~\ref{halpha_beta} we plot the effective temperatures obtained through the iron excitation/ionization equilibrium against those derived by the synthesis of the H$\alpha$ and H$\beta$ line profiles. No clear (and strong) systematic difference is seen
between the T$_{eff}$ derived by the two methods in the
temperature range covered by the stars studied in this paper. It is, therefore, wise to claim that the relation seen in Fig~\ref{halpha_beta} can be extended to the whole 63 stars considered in this study.\\
\indent As a conclusion we may assume that the temperatures derived using the analysis of the FeI and FeII lines agree reasonably well with those coming from the fitting of the H-line profiles (with the exception of the star CD-361785) even though in the case of the latter method another set of model atmospheres and another line formation code was used. A small systematic temperature difference may be present in the data, though not higher than $\sim$100\,K peek to peek. Such difference could account for no more than $\sim$0.06\,dex in the derived iron abundances (propagating the error in the temperature), and cannot thus fully explain (by a factor of 2) the [Fe/H] trend observed in Fig 2.

\indent We should note, however, that we are considering the effective temperatures derived using H-line profiles as reference. These values are likely also not free from systematic errors, that could arise from e.g. higher difficulties to place the continuum for the lower temperature stars, or from the filling-in of the profile for the very active stars. \\
\subsection{Astrometric Gravities}
\indent In addition to the test performed to the effective temperatures, we checked if there was any particular difference between our stellar gravities and those obtained through equation 1 in Santos et al. (\cite{sa04b}), using Hipparcos parallaxes (ESA \cite{esahip97}), hence accurate luminosities and T$_{eff}$, which enable mass estimation using evolutionary tracks.\\
\indent We should first stress that any conclusion yielded by this comparison should be taken with caution. The variability of the PTTS analysed, the uncertainties attached to PMS evolutionary tracks used for mass estimation are amongst some of the reasons why we must be careful in addressing stellar parameters coming from evolutionary stellar models. Indeed, stars in this phase of evolution are experiencing considerable structure contraction which can drive radius, age and mass estimations completely off the course. At the same time, Hipparcos paralaxes are not always sufficient to perform a whole sample comparison (in some cases the uncertainties are quite high).\\
\begin{figure}[t]
\centering
\includegraphics[scale=0.45]{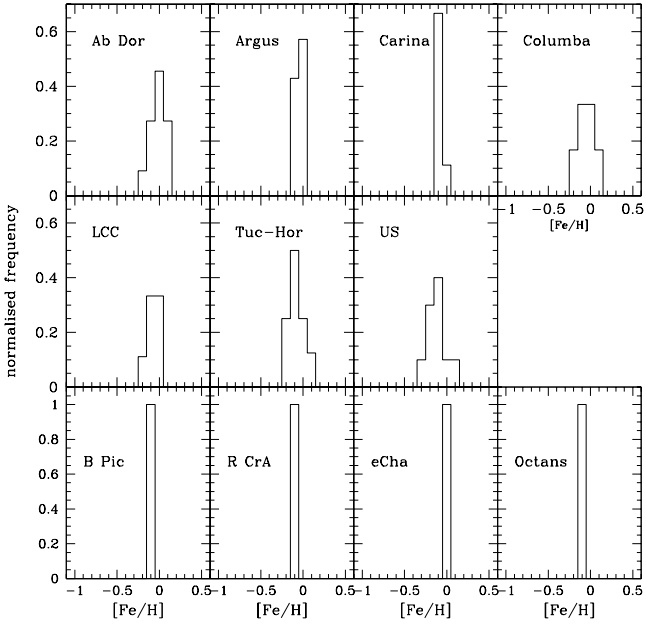}
\caption{Weighted distribution of our PTTS sample as a function of [Fe/H] for each association studied.}
\label{fig1}
\end{figure}
\indent Notwithstanding these considerations, we searched for parallaxe ($\pi$) information in the Hipparcos catalogue to compute stellar astrometric masses that could be used for the gravity comparison. Unfortunately, from the 63 stars of Table~\ref{kstars} only 19 featured the catalogue. A whole sample comparison between the gravities achieved through the two different techniques was therefore not feasible. \\
\indent Regardless of this, we estimated the mass for these 19 stars by using the isochrones of Shaller et al. (\cite{scha92}) and Shaerer et al. (\cite{sch92}, \cite{sch93}) when interpolating the V absolute magnitudes of Table~\ref{radec} with the T$_{eff}$ of Table~\ref{kstars} in the resulting Hertzprung-Russel diagram. For the Bolometric Correction we used the calibration by Flower (\cite{flower96}).\\
\indent A comparison of the two sets of surface gravities has shown that the parallax based values are, on average, higher than the ones derived through the use of FeI and FeII lines. The average offset is of about $\sim$ 0.46\,dex, with an rms of 0.32\,dex. The values derived using Hipparcos gravities for the 19 stars (average of 4.80\,dex) are clearly strongly above the expectation, while those derived by the method described in Sect.\,3.1 (average of 4.33\,dex) are usually within the expected range of values. \\
%
%
\section{Results}
\subsection{Fe abundance and atmospheric parameters}
\begin{figure}[t]
\centering
\includegraphics[scale=0.45]{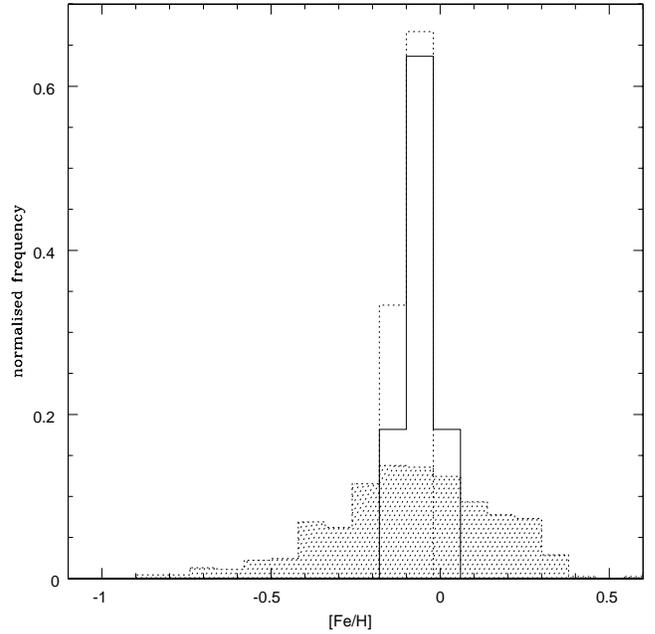}
\caption{Metallicity weighted distribution of the 6 SFR (dotted line) studied by SA08, of our 11 associations (solid line) and of the 450 MS stars (shadowed area) in the solar neighborhood plotted as functions of [Fe/H].}
\label{histo_med}
\end{figure}

By analysing the metallicity values of the different stars and the relatively small
dispersions obtained (Table~\ref{kstars}) we see that the trend described in section 3.2
rather than limiting the conclusions, it  strengthens the notion that the metallicities of
all associations studied here have approximately the same value, namely, around solar 
(Fig~\ref{fig1}). The typical standard deviations are within the uncertainty range of the
[Fe/H] determinations. \\
\indent In Figure~\ref{histo_med} the normalized distributions of the average
values of [Fe/H] for the 11 associations of our sample, and for two comparison samples from SA08
and SOU08 are shown. As we can seen in this Figure and also in Table~\ref{tablesfr}, low dispersion values (at the level of individual errors) 
are also found if the  dispersion of the average metallicity of all 11 associations
studied in this paper and the SFRs analysed by SA08 are considered.\\
\indent A striking feature of Figure~\ref{histo_med} is that while the MS population show a broad distribution of metallicities spanning
over more than 1.0 dex, the PMS sample of SA08 and the nearby young associations overlap almost entirely. Both distributions could be
combined without increasing the dispersion in the final [Fe/H] distribution. \\
\begin{figure*}[t!]
\includegraphics[scale=0.45]{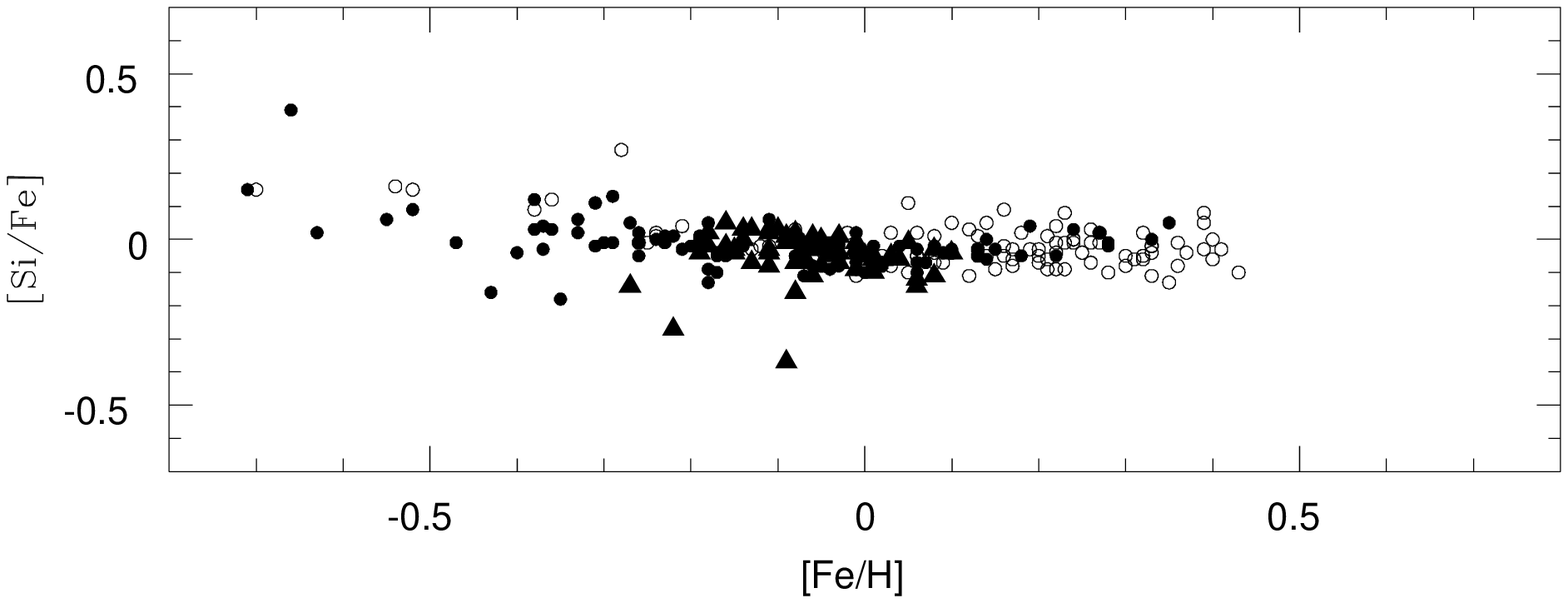}
\includegraphics[scale=0.45]{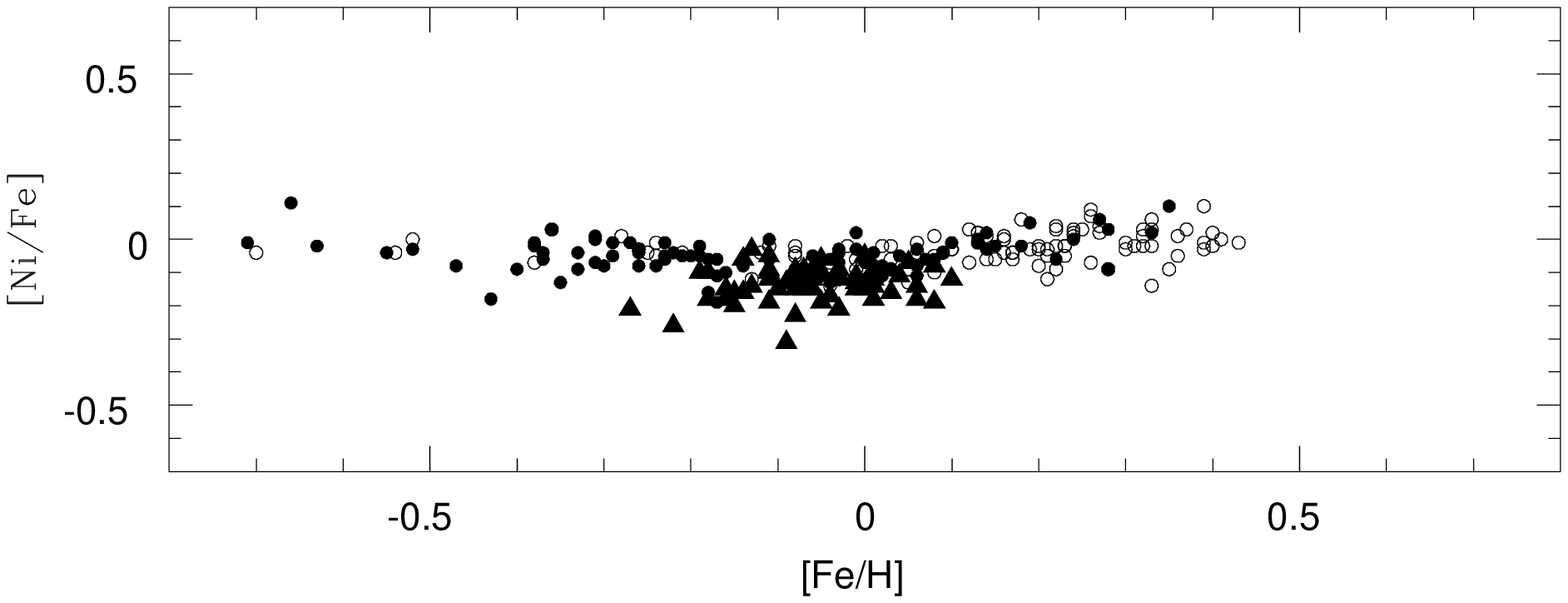}
\vspace*{-5.0cm}
\caption{[X/Fe] vs [Fe/H] for the PTTS in the sample (filled triangles) and for two comparison sets of stars with (open circles) and without planetary-mass companions (filled circles) in the solar neighborhood from the literature.}
\label{fig3}
\end{figure*}
\indent In summary, {\it the nearby young associations and the SFRs share the same metallicity. Within the errors, both show solar metallicities
with very little (below 0.1 dex) dispersion around the mean [Fe/H] value}. Although the exact formation mechanism of these 
young associations is not known, our result is another evidence suggesting that they are somehow all related to the 
nearby SFRs (Torres et al.~\cite{tor08}). Our results brings further support to the idea suggested by SA08 that super metal-rich stars
(with metallicity clearly above solar) found in the solar vicinity may have been formed in inner Galactic disk regions.\\
\setlength\extrarowheight{2pt}\vspace*{0.5pt}
\begin{table}[b]
\caption{Weighted average of the Fe abundance and standard deviation for all the associations in our sample (Total) and for the sample of SA08 before and after correction (asterisked columns).}             
\label{tablesfr}      
\centering                          
\begin{tabular}{cccccc}        
\hline\hline                 
Sample    & $\langle$[Fe/H]$\rangle$    &   $\sigma$  & $\langle$[Fe/H]$\rangle$* & $\sigma$* & N$_{SFR}$ \\  
\hline                                                 
\hline 
Total & -0.06 & 0.04 & 0.00 & 0.03 & 11\\
SA08 & -0.08 & 0.03 & 0.06 & 0.03 & 6\\
\hline 
\end{tabular}
\end{table}
\subsection{Other elements}
Several abundance surveys across the Galaxy over the years have shown that 
structural components of the Milky-Way (bulge, thin and thick disk and halo) 
present characteristic abundance patterns for $\alpha$-elements and iron-peak elements
with respect to Iron and Oxygen (see \textit{e.g.} McWilliam \cite{mac97}; Nissen
et al. \cite{nisse00}; Ortega et al. \cite{horto02}; Fullbright \cite{fulbri02}; Bensby et al. \cite{beni03}).\\
\indent In this sense, we have derived nickel (iron-peak element) and silicium ($\alpha$-element) abundances for the stars studied in this paper.
As in SA08, we choose these two elements because the list of spectral lines used is not considerably affected by non-LTE
effects. The derived abundances are given  in Table~\ref{elements}. Additionally, 
the average abundance [X/H] of all the sample along with the standard deviations and the average number of spectral lines used for
each species within each studied association is also given in Table~\ref{table:144}. \\
\indent In Fig.~\ref{fig3} we compare our results with those from GI06 for a sample of stars with and without planetary-mass companions in
the solar neighborhood.  Because Fe is believed to be a good indicator of the evolutionary state of an association it is extremely useful to plot [X/Fe] against [Fe/H]
as it sets a plane where we can compare our results with previous studies taking this element as a reference. 
>From Fig.~\ref{fig3} we can see that the abundance ratios of Si and Ni agree reasonably well with the ratios displayed by the comparison samples.
In fact, in the range of [X/Fe] ratios covered by our sample, there is almost an entire overlap of the data points of the two studies.\\
\addtocounter{table}{1} 
\setlength\extrarowheight{2pt}\vspace*{0.5pt}
\begin{table}[bt]
\caption{Average [X/H] abundance, corresponding standard deviation and average number of lines used for abundances determination.}             
\label{table:144}      
\centering                          
\begin{tabular}{c c c c}        
\hline\hline                 

Species   &	$\langle$[X/H]$\rangle$    & $\sigma$ & $\langle$N$\rangle$ \\  
\hline                                                 
Si        &	-0.11	   &	0.1 & 8.4 \\ 
Ni        & 	0.01	   &	0.10 & 23.3\\
\hline 
\end{tabular}
\end{table}
\indent In contrast to the Si, the nickel abundances disagree slightly from the main trend shown by the comparison sample. Our results are marginally
lower than the ones obtained for MS stars by GI06. At this point, we cannot find an explanation for this small discrepancy. It might reflect simply
small systematic errors in the analysis.\\
%
%
\indent As concluded by SA08 for the nearby SFRs, our results show that no clear differences exist between the
abundance ratios of the studied associations and those of field stars of
the same metallicity. The chemical abundances for the combined young sample (SFRs from SA08 and the young associations of this paper)
seem to reflect those of typical thin disk stars.
\section{Conclusions}
In this work, we have determined precise abundances for iron, silicium and nickel for 63 stars belonging to nearby young associations. 
While the main aim of this study was to search for a high metallicity sample within these associations, we were unable to detect any metal over-abundance in 
the combined sample of young stars (from this paper and those from SA08) where
[Fe/H] has been derived. \\
\indent Because both
groups of WTTS and PTTS of the two samples have not experienced significant spatial displacements, our results imply that
the high-metallicity stars harboring planets were not formed in the solar neighborhood
but were instead carried to their present position by the galactic dynamical forces. \\
\indent The Ni and Si abundances derived here are compatible with the general trend observed for the thin disk stellar population.\\
%
%
%

%
%
\begin{acknowledgements}
PVA and JFG were financially supported by the FCT project PTDC/CTE-AST/65971/2006. NCS would like to thank the support from Funda\c{c}\~ao para a
Ci\^encia e a Tecnologia, Portugal through programme Ci\^encia\, 2007 (C2007-CAUP-FCT/136/2006). MAvE is supported by an individual fellowship
(reference SFRH/BPD/26817/2006) granted by the Funda\c{c}\~ao para a Ci\^{e}ncia e a Tecnologia (FCT), Portugal. MAvE would like to thank Klaus Fuhrmann who supplied software and model atmospheres for the analysis of the Balmer lines. Furthermore Andreas Korn is thanked for providing a program for the co-addition of spectra.
\end{acknowledgements}

\setlength\extrarowheight{4pt}\vspace{0.5pt}
\longtab{1}{
\begin{longtable}{cccccc}
\caption{\label{radec} Stellar objects used in this study, their coordinates, V magnitude and the projected rotational velocity $\upsilon$ sin $i$. High probability members proposed for the different associations are identified with an H in the last column. The remaining stars are possible members (P).}\\
\hline\hline
Star &  RA  &  DEC  & V &  $\upsilon$ sin $i$& P.\\
&($\alpha_{2000}$)&($\delta_{2000}$)& (mag)& [km s$^{-1}$]&\\
\hline
\endfirsthead
\hline\hline
Star &  RA  &  DEC  & V & $\upsilon$ sin $i$ & P.\\
& ($\alpha_{2000}$) & ($\delta_{2000}$) & (mag) & [km s$^{-1}$]&\\
\hline
\endhead
\hline
\endfoot
\hline
\endlastfoot
& AB Dor & \hspace*{-20pt} Association & & &\\
%
CD-12 243      & 01 20 32.3 &-11 28 04 &  8.43  &  3  & H \\
CD-40 1701     & 05 02 30.4 &-39 59 13 & 10.57	&  6  & H \\
HD 37572       & 05 36 56.9 &-47 57 53 &  7.84	&  9  & H \\
HD 37551A      & 05 37 12.9 &-42 42 56 &  9.55	&  5  & H \\
HD 37551B      & 05 37 13.2 &-42 42 57 & 10.65	&  5  & H \\
CD-34 2676     & 06 08 33.9 &-34 02 55 & 10.17	&  13 & H \\
CD-84 80       & 07 30 59.5 &-84 19 28 &  9.96	&  4  & H \\
HD 64982       & 07 45 35.6 &-79 40 08 &  8.96	&  14 & H \\
TYC 8243 2975 1& 12 30 29.6 &-52 22 27 & 12.04	&  5  & P \\
HD 207278      & 21 48 48.5 &-39 29 09 &  9.66	& 9   & H \\
HD 217343      & 23 00 19.3 &-26 09 14 &  7.49	& 12  & H \\
HD 218860A     & 23 11 52.1 &-45 08 11 &  8.75	& 6   & H \\
\hline
& Argus & \hspace*{-20pt} Association & & &\\
%
CD-29 2360     & 05 34 59.2 &-29 54 04 & 10.64	&  11  & P  \\
CD-28 3434     & 06 49 45.4 &-28 59 17 & 10.38	&  6   & H \\
CD-42 2906     & 07 01 53.4 &-42 27 56 & 10.61	&  10  & H \\
TYC 8561 0970 1& 07 53 55.5 &-57 10 07 & 11.50	&   5  & H \\
BD-20 2977     & 09 39 51.4 &-21 34 17 & 10.22	&  10  & H \\
CD-39 5833     & 09 47 19.9 &-40 03 10 & 10.89	&  10  & H \\
CD-52 10232    & 22 39 30.3 &-52 05 17 & 10.85	&   8  & H \\
\hline
& $\beta$ Pic & \hspace*{-20pt} Association & & &\\
%
HD 322990     &  17 16 07.7 & -37 28 27 & 11.46	& 10   & P \\
\hline
& Carina & \hspace*{-20pt} Association & & &\\
%
TYC 8862 0019 1& 02 58 04.0 &-62 41 14 & 11.67	&  7   & P \\
HD 44627       & 06 19 12.9 &-58 03 16 &  9.13	& 11   & H \\
TYC 9178 0284 1& 06 55 25.2 &-68 06 21 & 11.91	& 11   & P \\
HD 55279       & 07 00 30.5 &-79 41 46 & 10.11	&  9   & H \\
CD-57 1709     & 07 21 23.7 &-57 20 37 & 10.72	& 11   & H \\
CD-55 2543     & 09 09 29.4 &-55 38 27 & 10.20	& 13   & H \\
HD 298936      & 10 13 14.8 &-52 30 54 &  9.79	&  9   & P \\
\hline
& Columba & \hspace*{-20pt} Association & & &\\
%
HD 26980       & 04 14 22.6 &-38 19 02 &  9.08	& 13   & H \\
HD 27679       & 04 21 10.3 &-24 32 21 &  9.43	& 11   & H\\
CD-36 1785     & 04 34 50.8 &-35 47 21 & 10.84	&  8   & H\\
HD 32372       & 05 00 51.9 &-41 01 07 &  9.50	&  7   & H\\
HD 274561      & 05 28 55.1 &-45 34 58 & 11.45	&  7   & H\\
CD-40 2458     & 06 26 06.9 &-41 02 54 & 10.00	& 12   & H\\
\hline
& R CrA & \hspace*{-20pt} Association & & &\\
%
CD-37 12759    & 18 39 05.3*&-37 26 22 & 10.98	& 9    & H\\
CD-36 13163    & 18 57 20.8*&-36 43 01 & 11.04	& 10   & H\\
\\
\hline
& $\epsilon$ Cha & \hspace*{-20pt} Association & & &\\
%
HD 105923      & 12 11 38.1 &-71 10 36 &  9.16	& 12   & H \\
\hline
& LCC & \hspace*{-20pt} Association & & &\\
%
CP-52 5025     & 11 55 57.7 &-52 54 01 & 11.00	&  6   & H\\
CD-49 4947     & 12 12 11.2 &-49 50 08 & 11.37	& 14   & H\\
CP-64 1859     & 12 19 21.6 &-64 54 10 &  9.87	& 13   & P \\
CD-51 6900     & 12 40 46.6 &-52 11 05 & 11.91	& 14   & H\\
CD-40 7581     & 12 56 12.3 &-41 22 20 & 11.73	& 13   & H\\
CD-40 8031     & 13 37 57.3 &-41 34 42 & 10.12	& 13   & H\\
CP-66 2366     & 13 54 07.4 &-67 33 45 & 10.93	& 11   & P \\
\hline
& Octans & \hspace*{-20pt} Association & & &\\
%
HD 23208       & 03 42 39.8 &-20 32 44 &  9.16	& 10   & P \\
\hline
& Tuc-Hor & \hspace*{-20pt} Association & & &\\ 
%
HD 105        &  00 05 52.5 &-41 45 11 &  7.53	& 12   &  H \\
HD 987        &  00 13 53.0 &-74 41 18 &  8.78	&  7   &  H\\
HD 8558       &  01 23 21.3 &-57 28 51 &  8.51	& 14   &  H\\
HD 9054       &  01 28 08.7 &-52 38 19 &  9.07	&  4   &  H\\
CD-46 1064    &  03 30 49.1 &-45 55 57 &  9.55	& 10   &  H\\
HD 47875      &  06 34 41.0 &-69 53 06 &  9.17	& 11   &  P\\
CD-38 4458    &  08 26 10.0 &-39 02 05 & 10.31	&  9   &  P \\
HD 202917     &  21 20 50.0 &-53 02 03 &  8.69	& 14   &  H\\
HD 222259B    &  23 39 39.3 &-69 11 40 &  9.84	& 14   &  H\\
\hline
& US & \hspace*{-20pt} Association & & & \\
%
CD-34 10180    & 15 07 14.8 &-35 05 00 & 10.53	& 13   & H\\
CD-36 10208    & 15 29 47.3 &-36 28 37 & 11.16	& 14   & H\\
TYC 9034 0968 1& 15 33 27.5 &-66 51 25 & 10.99	&  5   & H\\
CD-39 10162    & 15 47 41.8 &-40 18 27 & 11.08	& 12   & H\\
CD-25 11330    & 16 05 50.6 &-25 33 14 & 10.93	&  8   & H\\
CD-22 11502    & 16 19 34.0 &-22 28 29 & 11.11	& 12   & H\\
CD-51 10295    & 16 33 50.4 &-51 19 01 & 10.75	&  4   & H\\
CD-31 13486    & 17 02 27.8 &-32 04 36 & 10.13	& 13   & H\\
CD-23 13281    & 17 16 18.1 &-23 10 47 & 10.97	& 14   & H\\
TYC7886 1894 1 & 17 58 31.5 &-37 43 03 & 11.62	&  5   & H\\
\end{longtable}
}

\setlength\extrarowheight{4pt}\vspace{0.5pt}
\longtab{2}{
\begin{longtable}{ccccccccc}
\caption{\label{kstars}Results from the spectroscopic analysis -
  The determined atmospheric parameters T$_{eff}$, log$g$, [Fe/H],
  $\xi_t$ for each star. In the last three columns are shown the number
  of Fe lines used in each determination, the average uncertainty
  on the abundances given by each line EW. [Fe/H]* is the Iron abundance corrected 
  according to the method explained in section 3.2.}\\
\hline\hline
Star & T$_{eff}$ & log$g$ & $\xi_t$ & [Fe/H] & [Fe/H]* & N& $\sigma$\\
& [K] & [cm s$^{-2}$] & [km s$^{-1}$] & & & (FeI , FeII)& (FeI , FeII)\\
\hline
\endfirsthead
\hline\hline
Star & T$_{eff}$ & log$g$ & $\xi_t$ & [Fe/H] & [Fe/H]* & N& $\sigma$\\
& [K] & [cm s$^{-2}$] & [km s$^{-1}$] & & & (FeI , FeII)&(FeI , FeII)\\
\hline
\endhead
\hline
\endfoot
\hline
\endlastfoot
& & & AB Dor &\hspace*{-20pt} Association & & & &\\
%
CD-12 243  & 5406$\pm$38 & 4.67$\pm$0.09 & 1.21$\pm$0.02 &  0.00$\pm$0.05 & 0.04  & 37 , 10 & 0.05 , 0.03 \\
CD-40 1701 & 4800$\pm$80 & 4.61$\pm$0.24 & 1.68$\pm$0.14 & -0.11$\pm$0.08 & 0.01  & 36 ,  8 & 0.08 , 0.10 \\
HD 37572   & 5251$\pm$43 & 4.45$\pm$0.11 & 1.82$\pm$0.07 & -0.11$\pm$0.06 & -0.05 & 34 ,  6 & 0.05 , 0.05 \\
HD 37551A  & 5684$\pm$32 & 4.52$\pm$0.07 & 1.46$\pm$0.05 &  0.05$\pm$0.04 & 0.06  & 33 , 12 & 0.04 , 0.03 \\
HD 37551B  & 5055$\pm$68 & 4.31$\pm$0.14 & 1.65$\pm$0.08 & -0.03$\pm$0.08 & 0.06  & 36 ,  8 & 0.07 , 0.06 \\
CD-34 2676 & 5571$\pm$86 & 4.47$\pm$0.20 & 2.08$\pm$0.14 & -0.01$\pm$0.10 & 0.01  & 32 ,  8 & 0.08 , 0.08 \\ 
CD-84 80   & 5381$\pm$45 & 4.53$\pm$0.16 & 1.53$\pm$0.07 & -0.05$\pm$0.05 & 0.00 & 38 , 12 & 0.05 , 0.07 \\
HD 64982   & 6106$\pm$72 & 4.14$\pm$0.30 & 1.53$\pm$0.13 &  0.10$\pm$0.09 & 0.05  & 32 , 12 & 0.07 , 0.12 \\
TYC 8243 2975 1 & 4626$\pm$74  & 4.16$\pm$0.22 & 1.35$\pm$0.11 & -0.19$\pm$0.07 & -0.04 & 31 , 6 & 0.07 , 0.09\\ 
HD 207278  & 5714$\pm$52 & 4.51$\pm$0.10 & 1.45$\pm$0.08 &  0.08$\pm$0.06 & 0.08  & 33 ,  9 & 0.05 , 0.05\\
HD 217343  & 5795$\pm$41 & 4.44$\pm$0.09 & 1.68$\pm$0.08 &  0.00$\pm$0.05 & -0.01 & 34 , 10 & 0.04 , 0.05\\
HD 218860A & 5550$\pm$68 & 4.50$\pm$0.13 & 1.42$\pm$0.05 &  0.10$\pm$0.06 & 0.13  & 34 , 12 & 0.05 , 0.05\\
\hline
& & & Argus &\hspace*{-30pt} Association & & & &\\
%
CD-29 2360  & 4917$\pm$77 & 4.66$\pm$0.16 & 1.92$\pm$0.12 &  0.04$\pm$0.07 & 0.15  & 32 ,  5 & 0.08 , 0.06\\
CD-28 3434  & 5652$\pm$39 & 4.51$\pm$0.13 & 1.62$\pm$0.07 & -0.01$\pm$0.05 & 0.00  & 35 , 11 & 0.04 , 0.06 \\ 
CD-42 2906  & 5308$\pm$49 & 4.38$\pm$0.13 & 1.76$\pm$0.07 & -0.03$\pm$0.07 & 0.03  & 36 ,  8 & 0.06 , 0.06 \\
TYC 8561 0970 1  & 5348$\pm$51 & 4.48$\pm$0.15 & 1.75$\pm$0.07 & -0.07$\pm$0.07 & -0.02 & 38 , 12 & 0.06 , 0.06\\
BD-20 2977  & 5421$\pm$50 & 4.50$\pm$0.14 & 2.05$\pm$0.09 & -0.09$\pm$0.06 & -0.05 & 33 , 10 & 0.05 , 0.07\\
CD-39 5833  & 5471$\pm$61 & 4.45$\pm$0.12 & 1.72$\pm$0.10 &  0.01$\pm$0.07 & 0.05  & 35 ,  9 & 0.07 , 0.05\\
CD-52 10232 & 5381$\pm$59 & 4.48$\pm$0.13 & 1.83$\pm$0.09 & -0.06$\pm$0.07 & -0.01 & 35 , 10 & 0.07 , 0.06\\ 
\hline
 & & & $\beta$ Pic &\hspace*{-30pt} Association & & & &\\
%
HD 322990   & 4858$\pm$74 & 4.20$\pm$0.20 & 2.05$\pm$0.10 & -0.13$\pm$0.08 & -0.01 & 31 , 9 & 0.08 , 0.10\\
\hline
 & & & Carina &\hspace*{-20pt} Association & & & & \\
%
TYC 8862 0019 1 & 4776$\pm$120 & 4.38$\pm$0.47 & 2.29$\pm$0.2  & -0.14$\pm$0.14 & -0.01 & 38 , 10 & 0.14 , 0.25\\
HD 44627   & 5156$\pm$54  & 4.29$\pm$0.17 & 1.90$\pm$0.06 & -0.01$\pm$0.06 & 0.07  & 30 ,  8 & 0.05 , 0.08 \\
TYC 9178 0284 1 & 4607$\pm$89  & 3.98$\pm$0.36 & 1.90$\pm$0.12 & -0.10$\pm$0.08 & 0.05  & 28 ,  6 & 0.08 , 0.19\\
HD 55279   & 4918$\pm$50  & 4.19$\pm$0.18 & 1.85$\pm$0.07 & -0.08$\pm$0.05 & 0.03  & 33 ,  8 & 0.05 , 0.08 \\
CD-57 1709 & 5368$\pm$55  & 4.58$\pm$0.19 & 1.85$\pm$0.08 & -0.06$\pm$0.06 & -0.01 & 33 , 10 & 0.05 , 0.09 \\ 
CD-55 2543 & 5512$\pm$53  & 4.38$\pm$0.20 & 1.70$\pm$0.08 & -0.05$\pm$0.06 & -0.02 & 31 ,  8 & 0.05 , 0.09 \\
HD 298936  & 4923$\pm$119 & 4.47$\pm$0.29 & 1.96$\pm$0.14 & -0.08$\pm$0.08 & 0.03  & 30 , 8 & 0.06 , 0.08 \\
%
\hline
 & & & Columba &\hspace*{-20pt} Association & & & & \\
%
HD 26980   & 5910$\pm$61 & 4.26$\pm$0.12 & 1.74$\pm$0.12 & -0.01$\pm$0.07 & -0.03 & 34 , 10 & 0.07 , 0.06 \\
HD 27679   & 5817$\pm$58 & 4.16$\pm$0.31 & 1.96$\pm$0.13 & -0.04$\pm$0.08 & -0.05 & 37 , 12 & 0.07 , 0.14 \\
CD-36 1785 & 5276$\pm$48 & 4.31$\pm$0.19 & 2.63$\pm$0.09 & -0.15$\pm$0.06 & -0.09 & 32 ,  7 & 0.05 , 0.09\\
HD 32372   & 5716$\pm$50 & 4.31$\pm$0.09 & 1.89$\pm$0.10 & -0.07$\pm$0.07 & -0.07 & 38 , 11 & 0.06 , 0.04\\
HD 274561  & 5144$\pm$63 & 4.36$\pm$0.12 & 2.09$\pm$0.10 & -0.11$\pm$0.07 & -0.03 & 30 ,  7 & 0.06 , 0.05\\
CD-40 2458 & 5415$\pm$62 & 4.32$\pm$0.19 & 1.98$\pm$0.09 &  0.06$\pm$0.08 & 0.1	 & 34 ,  9 & 0.07 , 0.09 \\
\hline
 & & & R CrA &\hspace*{-20pt} Association & & & &\\
%
CD-37 12759 & 5149$\pm$104 & 4.28$\pm$0.30 & 1.98$\pm$0.15 & -0.08$\pm$0.13 & 0.00 & 35 ,  10 & 0.07 , 0.07\\
CD-36 13163 & 5098$\pm$44  & 3.99$\pm$0.13 & 1.73$\pm$0.05 & -0.08$\pm$0.06 & 0.00 & 31 ,   9 & 0.05,0.07\\
\hline
 & & & $\epsilon$ Cha &\hspace*{-30pt} Association & & & & \\
%
HD 105923   & 4979$\pm$40  & 3.98$\pm$0.13 & 1.94$\pm$0.06 &  0.01$\pm$0.05 & 0.03  & 32 , 8 & 0.05 , 0.06 \\
\hline
& & & LCC & \hspace*{-20pt} Association & & & &\\
%
CP-52 5025 & 4834$\pm$78  & 4.19$\pm$0.18 & 1.69$\pm$0.10 & -0.07$\pm$0.07 & 0.05  & 28 , 7 & 0.07 , 0.06 \\
CD-49 4947 & 4875$\pm$56  & 4.17$\pm$0.20 & 2.21$\pm$0.07 & -0.08$\pm$0.06 & 0.03  & 28 , 6 & 0.06 , 0.10 \\
CP-64 1859 & 5422$\pm$145 & 5.14$\pm$0.25 & 1.46$\pm$0.25 & -0.08$\pm$0.13 & 0.03 & 30 , 7 & 0.11 , 0.09 \\ 
CD-51 6900 & 4842$\pm$71  & 3.98$\pm$0.18 & 2.04$\pm$0.10 & -0.15$\pm$0.08 & -0.03 & 29 , 4 & 0.07 , 0.08\\
CD-40 7581 & 4979$\pm$69  & 4.20$\pm$0.13 & 1.80$\pm$0.08 &  0.01$\pm$0.07 & 0.11  & 30 , 4 & 0.07 , 0.04\\
CD-40 8031 & 5243$\pm$55  & 4.06$\pm$0.21 & 1.78$\pm$0.07 & -0.03$\pm$0.07 & 0.04  & 31 , 7 & 0.06 , 0.10\\ 
CP-66 2366 & 5793$\pm$75  & 4.89$\pm$0.34 & 2.15$\pm$0.15 & -0.03$\pm$0.08 & -0.04 & 25 , 9 & 0.06 , 0.15\\
\hline
& & & Octans & \hspace*{-20pt} Association & & & & \\
%
HD 23208  & 5322$\pm$43 & 4.23$\pm$0.14 & 1.85$\pm$0.06 & -0.09$\pm$0.06 & -0.03 & 34 , 8 & 0.05 , 0.07\\
\hline
& & & Tuc-Hor & \hspace*{-20pt} Association & & & & \\ 
%
HD 105	  & 6012$\pm$68 & 4.42$\pm$0.12 & 1.24$\pm$0.12 &  0.06$\pm$0.09 & 0.02  & 36 ,  8 & 0.07 , 0.05\\
HD 987	  & 5488$\pm$46 & 4.36$\pm$0.12 & 1.85$\pm$0.08 & -0.07$\pm$0.06 & -0.04 & 35 , 11 & 0.05 , 0.06 \\
HD 8558    & 5538$\pm$53 & 4.04$\pm$0.16 & 1.90$\pm$0.09 & -0.09$\pm$0.07 & -0.06 & 30 ,  7 & 0.06 , 0.07 \\ 
HD 9054    & 5045$\pm$55 & 4.49$\pm$0.15 & 1.90$\pm$0.07 & -0.08$\pm$0.06 & 0.01  & 31 ,  9 & 0.05 , 0.07 \\ 
CD-46 1064 & 4777$\pm$70 & 4.13$\pm$0.18 & 1.92$\pm$0.10 & -0.16$\pm$0.07 & -0.03 & 33 ,  6 & 0.07 , 0.07 \\
HD 47875   & 5781$\pm$35 & 4.53$\pm$0.17 & 1.75$\pm$0.07 &  0.01$\pm$0.05 & 0.00 & 33 ,  9 & 0.04 , 0.07\\
CD-38 4458 & 5751$\pm$34  & 4.32$\pm$0.11 & 1.60$\pm$0.06 &  0.03$\pm$0.05 & 0.03  & 35 , 10 & 0.04 , 0.05\\ 
HD 202917  & 5592$\pm$79 & 4.31$\pm$0.17 & 2.16$\pm$0.15 & -0.06$\pm$0.10 & -0.04 & 33 ,  9 & 0.08 , 0.07\\
HD 222259B & 4938$\pm$91 & 4.12$\pm$0.26 & 2.59$\pm$0.15 & -0.22$\pm$0.09 & -0.11 & 27 ,  8 & 0.08 , 0.12\\
\hline
 & & & US & \hspace*{-30pt} Association & & & & \\
%
CD-34 10180 & 5188$\pm$49 & 4.05$\pm$0.20 & 2.30$\pm$0.08 & -0.11$\pm$0.07 & -0.04 & 34 ,  8 & 0.06 , 0.10\\
CD-36 10208 & 4869$\pm$71 & 3.64$\pm$0.23 & 2.59$\pm$0.10 & -0.27$\pm$0.07 & -0.16 & 27 ,  5 & 0.07 , 0.11\\
TYC 9034 0968 1  & 4955$\pm$65 & 4.52$\pm$0.47 & 1.80$\pm$0.07 & -0.11$\pm$0.11 & -0.01 & 33 ,  9 & 0.07 , 0.13\\
CD-39 10162 & 4922$\pm$61 & 4.05$\pm$0.14 & 2.04$\pm$0.09 & -0.14$\pm$0.08 & -0.03 & 33 ,  7 & 0.07 , 0.10\\
CD-25 11330 & 5076$\pm$62 & 4.14$\pm$0.18 & 2.02$\pm$0.09 & -0.16$\pm$0.07 & -0.07 & 34 , 11 & 0.06 , 0.08 \\
CD-22 11502 & 4951$\pm$66 & 3.81$\pm$0.17 & 2.10$\pm$0.08 & -0.18$\pm$0.07 & -0.08 & 34 ,  9 & 0.07 , 0.08\\
CD-51 10295 & 4922$\pm$58 & 4.31$\pm$0.20 & 1.59$\pm$0.07 & -0.08$\pm$0.05 & 0.03  & 31 ,  6 & 0.06 , 0.10 \\
CD-31 13486 & 5625$\pm$42 & 4.55$\pm$0.25 & 1.68$\pm$0.07 &  0.01$\pm$0.06 & 0.03  & 35 , 10 & 0.06 , 0.11\\
CD-23 13281 & 4900$\pm$84 & 4.41$\pm$0.26 & 2.14$\pm$0.32 &  0.08$\pm$0.17 & 0.19  & 28 ,  6 & 0.08 , 0.10\\
TYC 7886 1894 1  & 5048$\pm$62 & 4.26$\pm$0.18 & 2.16$\pm$0.09 & -0.18$\pm$0.07 & -0.09 & 33 ,  9 & 0.06 , 0.08\\
\end{longtable}
}

\setlength\extrarowheight{4pt}
\longtab{5}{
\setlength\tabcolsep{4pt}
\begin{longtable}{ccccccc}
\caption{\label{elements} Derived abundances of Si and Ni for the stars in the sample.}\\
\hline\hline
Star & [Si/H] & $\sigma$ & \textit{N} & [Ni/H] & $\sigma$ & \textit{N}\\
\hline
\endfirsthead
\hline\hline
Star & [Si/H] & $\sigma$ & \textit{N} & [Ni/H] & $\sigma$ & \textit{N}\\ 
\hline
\endhead
\hline
\endfoot
\hline
\endlastfoot
& & AB Dor & & & &\\
%
CD-12 243    &  -0.05  & 0.03 & 9 & -0.05  & 0.04 & 25\\
CD-40 1701   &  -0.08  & 0.10 & 7 & -0.16  & 0.10 & 24\\
HD 37572     &  -0.09  & 0.04 & 9 & -0.02  & 0.07 & 25\\
HD 37551A    &  0.04   & 0.04 & 9 & 0.02   & 0.06 & 30\\
HD 37551B    &  -0.02  & 0.04 & 9 & 0.02   & 0.07 & 27\\
CD-34 2676   &  -0.14  & 0.09 & 9 & -0.06  & 0.09 & 20\\
CD-84 80     &  -0.05  & 0.04 & 9 & -0.04  & 0.05 & 25\\
HD 64982     &  0.02   & 0.07 & 9 & -0.26  & 0.14 & 22\\
TYC 8243 2975 1 & -0.23  & 0.08 & 6 & -0.10  & 0.18 & 24\\ 
HD 207278    &  0.05   & 0.03 & 9 & 0.13   & 0.08 & 27\\
HD 217343    &  -0.05  & 0.05 & 9 & 0.07   & 0.06 & 22\\
HD 218860A   &  0.06   & 0.03 & 9 & 0.1    & 0.07 & 29\\
%
\hline
& & Argus & & & &\\
%
CD-29 2360   &  -0.02  & 0.09 & 9 & -0.03  & 0.08 & 20\\
CD-28 3434   &  -0.04  & 0.04 & 9 & -0.01  & 0.06 & 28\\
CD-42 2906   &  -0.07  & 0.07 & 9 & 0.06   & 0.06 & 26\\
TYC 8561 0970 1 & -0.09  & 0.04 & 9 & -0.01  & 0.06 & 27\\
BD-20 2977   &  -0.1   & 0.06 & 9 & -0.01  & 0.07 & 22\\
CD-52 10232  &  -0.05  & 0.03 & 9 & 0.04   & 0.08 & 26\\
CD-39 5833   &  -0.07  & 0.05 & 9 & 0.08   & 0.07 & 27\\
\hline
 & & $\beta$ Pic & & & &\\
%
HD 322990    &  -0.1   & 0.13 & 9 & 0.02   & 0.09 & 23\\
\hline
 & & Carina & & & &\\
%
TYC 8862 0019 1 & -0.11  & 0.13 & 8 & -0.10  & 0.12 & 21\\ 
HD 44627     &  -0.02  & 0.10 & 9 & 0.10   & 0.09 & 24\\
TYC 9178 0284 1 & -0.07  & 0.12 & 8 & 0.05   & 0.11 & 19\\ 
HD 55279     &  -0.07  & 0.07 & 9 & -0.01  & 0.08 & 23\\ 
CD-57 1709   &  -0.09  & 0.07 & 8 & 0.01   & 0.08 & 24\\
CD-55 2543   &  -0.13  & 0.07 & 8 & 0.00   & 0.11 & 18\\ 
HD 298936     &  -0.08  & 0.11 & 9 & -0.05  & 0.09 & 22\\
%
\hline
& & Columba & & & & \\
%
HD 26980     &  -0.1   & 0.08 & 8 & -0.16  & 0.06 & 17\\
HD 27679     &  -0.09  & 0.06 & 8 & 0.11   & 0.10 & 21\\
CD-36 1785   &  -0.18  & 0.08 & 9 & 0.06   & 0.08 & 25\\
HD 32372     &  -0.12  & 0.04 & 9 & 0.06   & 0.06 & 27\\
HD 274561    &  -0.15  & 0.07 & 9 & 0.04   & 0.08 & 25\\
CD-40 2458   &  -0.08  & 0.08 & 7 & 0.15   & 0.09 & 24\\
\hline
& & R CrA & & & &\\
%
CD-37 12759  &  -0.09  & 0.08 & 9 & 0.04   & 0.08 & 22\\
CD-36 13163  &  -0.06  & 0.06 & 9 & 0.05   & 0.09 & 25\\
\hline
 & & $\epsilon$ Cha & & & & \\
%
HD 105923    &    -    &   -  & - & -     &  -   &  -\\
\hline
& & LCC & & & &\\
%
CP-52 5025   &  -0.14  & 0.10 & 8 & 0.10   & 0.09 & 23\\
CD-49 4947   &  -0.15  & 0.10 & 5 & -0.08  & 0.13 & 15\\
CP-64 1859   &  -0.24  & 0.18 & 7 & -0.13  & 0.13 & 17\\
CD-51 6900   &  -0.19  & 0.08 & 8 & 0.03   & 0.11 & 19\\
CD-40 7581   &  -0.09  & 0.11 & 9 & 0.07   & 0.12 & 24\\
CD-40 8031   &  -0.1   & 0.09 & 8 & 0.07   & 0.09 & 18\\
CP-66 2366   &  -0.10  & 0.07 & 7 & -0.04  & 0.06 & 15\\
\hline
& & Octans & & & &\\
%
HD 23208     &  -0.08  & 0.09 & 9 & 0.02 & 0.08 & 27\\
\hline
& & Tuc-Hor & & & &\\
%
HD 105       &  -0.06  & 0.07 & 9 & 0.16   & 0.17 & 20\\
HD 987       &  -0.12  & 0.04 & 9 & 0.06   & 0.09 & 28\\
HD 8558      &  -0.46  & 0.31 & 3 & -0.4   & 0.34 & 11\\
HD 9054      &  -0.09  & 0.08 & 9 & -0.06  & 0.08 & 23\\
CD-46 1064   &  -0.11  & 0.11 & 9 & -0.07  & 0.10 & 21\\
HD 47875     &  -0.05  & 0.06 & 9 & 0.14   & 0.11 & 22\\
CD-38 4458   &  -0.02  & 0.09 & 9 & 0.12   & 0.07 & 26\\
HD 202917    &  -0.17  & 0.06 & 8 & 0.05   & 0.17 & 23\\
HD 222259B   &  -0.49  & 0.14 & 6 & 0.00   & 0.17 & 19\\
\hline
& & US &  & & &\\
%
CD-34 10180  &  -0.19  & 0.09 & 8 & 0.1    & 0.12 & 22\\
CD-36 10208  &  -0.41  & 0.11 & 6 & 0.07   & 0.13 & 20\\
TYC 9034 0968 1 & -0.14  & 0.11 & 9 & -0.1   & 0.10 & 25\\
CD-39 10162  &  -0.14  & 0.09 & 9 & -0.03  & 0.10 & 24\\
CD-25 11330  &  -0.18  & 0.06 & 9 & -0.02  & 0.11 & 27\\
CD-22 11502  &  -0.20  & 0.07 & 8 & 0.04   & 0.11 & 23\\
CD-51 10295  &  -0.09  & 0.08 & 9 & 0.00   & 0.10 & 25\\
CD-31 13486  &  -0.04  & 0.04 & 9 & 0.07   & 0.09 & 25\\
CD-23 13281  &  -0.03  & 0.10 & 8 & 0.07   & 0.16 & 24\\
TYC 7886 1894 1 & -0.16  & 0.08 & 9 & 0.09   & 0.16 & 29\\
\end{longtable}
}


\begin{thebibliography}{}

    \bibitem[1989]{ag89} Anders. E., \& Grevesse, N. 1989, Geochim, Cosmochim. Acta, 53, 197

    \bibitem[2006]{ammler06} Ammler M. 2006, PhD thesis, university of Jena

    \bibitem[2008]{ammsant08} Ammler - von Eiff, M. \& Santos, N.C. 2008, AN, 329, 573

    \bibitem[1998]{barna98} Barrado y Navascu{\'e}s, D. 1998, Ap\&SS, 263, 235

    \bibitem[2003]{beni03} Bensby, T., Feltzing, S. \& Lundstr\"{o}m, I. 2003, A\&A, 410, 527

    \bibitem[2004]{claynitt04} Clayton, D.D., Nittler, L.R. 2004 in Origin and Evolution of the Elements, 
    Ed. McWilliam, A., Rauch M. (Cambridge University Press), 297
    
    \bibitem[2000]{dek00} Dekker, H., D' Odorico, S., Kaufer, A., Delabre, B., \& Kotzlowski, H. 2000, SPIE, 4008, 534

    \bibitem[1997]{esahip97} ESA 1997, The Hipparcos and Tycho Catalogues, ESA SP-1200
    
    \bibitem[1999]{femo99} Feigelson, E.D., {Montmerle}, T. 1999 ARAA, 37, 363

    \bibitem[2005]{fixe05} Fischer, D.\& Valenti, J. 2005, ApJ, 622, 1102

    \bibitem[1996]{flower96} Flower, P. J. 1996, ApJ, 469, 355

    \bibitem[2004]{frufru04} Fuhrmann, K. 2004, Astron. Nachrichten, 325, 3

    \bibitem[2002]{fulbri02} Fulbright, J. P. 2002, AJ, 123, 404

    \bibitem[2006]{gi06} Gilli, G., Israelian, G., Ecuvillon, A., Santos, N.C., Mayor, M. 2006, ApR, 449, 723

    \bibitem[1997]{go97} Gonzalez, G. 1997, MNRAS, 285, 403

    \bibitem[1998]{gon98} Gonzalez, G. 1998 A\&A, 334, 221

    \bibitem[2001]{go01} Gonzalez, G., Laws, C., Tyagi, S., \& Reddy, B. E. 2001, AJ, 121, 432

    \bibitem[1992]{gregorio92} Gregorio-Hetem, J., Lepine, J. R. D., Quast, G. R., Torres, C. A. O., de La Reza, R. 1992, AJ 103, 549 
	

    \bibitem[1978]{he78} Herbig, G.H. 1978 in Problems of Physics and Evolution of the Universe, ed. Mirzoyan L.V. (Erevan:Armenian 
    Acad. Sci.), 171

    \bibitem[2006]{ja06} James, D.J., Melo, C., Santos, N.C., Bouvier, J. 2006, A\&A, 446, 971

    \bibitem[2003]{Kaca03} {Kasting}, J.~F. and {Catling}, D. 2003, ARAA, 41, 429

    \bibitem[1997]{kas97} Kastner, J.H., Zuckerman, B., Weintraub, D.A., Forveille, T. 1997, \textit{Science},
     277, 67

    \bibitem[1993]{kur93} Kurucz, R. 1993, ATLAS9 Stellar Atmosphere Programs and 2 km/s grid. Kurucz CD-ROM No. 13, Cambridge, Mass.: 
    Smithsonian Astrophysical Observatory, 1993, 13

    \bibitem[1999]{mamaj99} Mamajek, E.E., Lawson, W.A., Feigelson, E.D. 1999, ApJ, 516, 77

    \bibitem[2000]{mamaj00} Mamajek, E.E., Lawson, W.A., Feigelson, E.D. 2000, ApJ, 544, 356

    \bibitem[1997]{mac97} McWilliam, A., 1997, ARA\&A, 35, 503
    
    \bibitem[2004]{morel04} Morel, T., Micela, G., F., \& Katz, D. 2004, A\&A, 426, 1007

    \bibitem[2000]{neu00}{Neuh{\"a}user}, R. and {Walter}, F.~M. and {Covino}, E. and {Alcal{\'a}}, J.~M. and {Wolk}, S.~J. and {Frink}, S. 
    and {Guillout}, P. and {Sterzik}, M.~F. and {Comer{\'o}n}, F. 2000 \aaps, 146, 323

    \bibitem[2000]{nisse00} Nissen, P.E., Chen, Y.Q., Shuster, W.J., \& Zhao, G. 2000, A\&A, 353, 722

    \bibitem[2002]{horto02} Ortega, V. G., de la Reza, R., Jilinski, E., Bazzanella 2002, ApJ, 575, L75

    \bibitem[1996]{padgee96} Padgett, D.L. 1996, ApJ, 471, 847

    \bibitem[1996]{pol96} Pollack, J.~B., Hubickyj, O., Bodenheimer, P., Lissauer, J.~J., Podolak, M., 
            Greenzweig, Y. 1996, Icarus 124, 62 

    \bibitem[2001]{sa01} Santos, N. C., Israelian, G., \& Mayor, M. 2001, A\&A, 373, 1019

    \bibitem[2004]{sa04b} Santos, N.C., Israelian, G., \& Mayor, M. 2004, A\&A, 415, 1153

    \bibitem[2005]{santo05} Santos, N. C., Israelian, G., Mayor, M., et al. 2005, A\&A, 437, 1127

    \bibitem[2006]{sant06} Santos, N.C., Ecuvillon, A., Israelian, G., et al. 2006, A\&A, 458, 997

    \bibitem[2008]{sa08} Santos, N.C., Melo, C., James, D.J., Gameiro, J.F., Bouvier, J., Gomes J.I. 2008, A\&A, 480, 889

    \bibitem[1992]{sch92} Schaerer, D., Meynet, G., Maeder, A., et al. 1992, A\&AS, 98, 523

    \bibitem[1992]{scha92} Schaller, G., Schaerer, D., Meynet, G., \& Maeder, A. 1992, A\&AS,96, 269

    \bibitem[1993]{sch93} Schaerer, D., Charbonnel, C., Meynet, G., et al. 1993, A\&AS, 102,339

    \bibitem[1973]{sne73} Sneden, C. 1973, PhD. Thesis, Univ. of Texas

    \bibitem[2003]{songa03} Song, I., Zuckermann, B., Bessel, M.S. 2003, ApJ, 599, 342

    \bibitem[2008]{sousinha} Sousa, S.G., Santos, N.C., Mayor, M., Udry, S., Casagrande, L., Israelian, G., Pepe, F., Queloz, D., 
     Monteiro, M.~J.~P.~F.~G 2008, A\&A, 487, 373

    \bibitem[2001]{spa01} Spangler, C., Sargent, A.I., Silverstone, M.D., Becklin, E.E., Zuckerman, B. 2001, 
    ApJ, 555, 932

    \bibitem[2000]{to00} Torres, C.A.O., Quast, G.R., de La Reza, R., Jilinsky, E. 2000, AJ, 120, 1410
 
    \bibitem[2003b]{tor03} Torres, C.A.O., Quast, G.R., de La Reza, R., 
	da Silva, L., Melo, C.H.F., Sterzik, M. 2003b in Open Issues in Local Star Formation, ed. L{\'e}pine, J. and Gregorio-Hetem, J. 
        (Kluwer Academic Publishers), 299, 83

    \bibitem[2006]{tor06} Torres, C.A.O., Quast, G.R., da Silva, L., 
	de La Reza, R., Melo, C.H.F., Sterzik, M. 2006, AAP, 460, 695

    \bibitem[2008]{tor08} Torres, C.A.O., Quast, G.R., Melo, C.H.F., Sterzik, M. 2008, [arXiv:astro-ph/0808.3362] 

    \bibitem[1996]{wiele96} Wielen, R., Fuchs, B., \& Dettbarn, C. 1996, A\&A, 314, 438

    \bibitem[2000]{zuwe00} Zuckermann, B., Webb, R.A. 2000, Apj, 535, 959

    \bibitem[2001a]{zucka01} Zuckermann, B., Song, I., Bessel, M.S., \& Webb, R.A. 2001a, ApJ, 562, L87 

    \bibitem[2004]{zucksong} Zuckerman, B., and Song, I. 2004, ARA\&A, 42, 685

    \bibitem[2004c]{zusobe04} Zuckerman, B., Song, I., Bessell, M.S. 2004c, ApJ, 613, L65

\end{thebibliography}
\end{document}